\documentclass[Physsubmission, Phys]{SciPost}

\hypersetup{
    colorlinks,
    linkcolor={red!50!black},
    citecolor={blue!50!black},
    urlcolor={blue!80!black}
}

\usepackage[bitstream-charter]{mathdesign}
\DeclareSymbolFont{usualmathcal}{OMS}{cmsy}{m}{n}
\DeclareSymbolFontAlphabet{\mathcal}{usualmathcal}

\begin{document}

\begin{center}{\Large \textbf{Some Recent Results on Renormalization-Group
      Properties of Quantum Field Theories}}\end{center}

\begin{center}
Robert Shrock
\end{center}

\begin{center}
  C. N. Yang Institute for Theoretical Physics and Department of
  Physics and Astronomy, Stony Brook University, Stony Brook NY 11794, USA, 
  robert.shrock@stonybrook.edu
\end{center}

\begin{center}
\today
\end{center}

\definecolor{palegray}{gray}{0.95}
\begin{center}
\colorbox{palegray}{
  \begin{minipage}{0.95\textwidth}
    \begin{center}
    {\it  XXXIII International (ONLINE) Workshop on High Energy Physics \\“Hard Problems of Hadron Physics:  Non-Perturbative QCD \& Related Quests”}\\
    {\it November 8-12, 2021} \\
    \doi{10.21468/SciPostPhysProc.?}\\
    \end{center}
  \end{minipage}
}
\end{center}

\section*{Abstract}
{\bf  We discuss some higher-loop studies of renormalization-group flows and
  fixed points in various quantum field theories.}

\vspace{10pt}
\noindent\rule{\textwidth}{1pt}
\tableofcontents\thispagestyle{fancy}
\noindent\rule{\textwidth}{1pt}
\vspace{10pt}

\section{Introduction}
\label{intro_section}

A fundamental question in quantum field theory (QFT) concerns how the
running coupling of a theory changes as a function of the reference
Euclidean energy/momentum scale $\mu$ where it is measured.  The
variation of this coupling with $\mu$ is described by the
renormalization group (RG) beta function of the theory. Here
we will discuss some results that we have 
obtained in this area.  Much of this work was with T. A. Ryttov.  We
will focus mainly on vectorial asymptotically free nonabelian gauge
theories in $d=4$ dimensions, but also discuss some other asymptotically
free theories, namely the 2D finite-$N$ Gross-Neveu model and 6D
$\phi^3$ theories, as well as some infrared-free theories, including
U(1) gauge theory, O($N$) $\phi^4$ theory, and chiral gauge theories..

\section{Asymptotically Free Nonabelian Gauge Theories}
\label{nagt_section}

Let us consider an asymptotically free (AF) vectorial nonabelian gauge theory
(in $d=4$ dimensions) with gauge group $G$ and $N_f$ massless fermions
$\psi_j$, $j=1,..., N_f$, transforming according to a representation
$R$ of $G$.  We denote the running gauge coupling as $g(\mu)$ and
define $\alpha(\mu) \equiv g(\mu)^2/(4\pi)$ and $a(\mu) \equiv
g(\mu)^2/(16\pi^2)$. The dependence of $\alpha(\mu)$ on $\mu$ is
described by the RG beta function, $\beta = d\alpha(\mu)/dt$, where
$dt=d\ln\mu$.  This has the series expansion
\begin{equation}
\beta = -2\alpha \sum_{\ell=1}^\infty b_\ell \, a^\ell \ , 
\label{beta}
\end{equation}
where $b_\ell$ is the $\ell$-loop coefficient.  For a general operator
${\cal O}$, we denote the full scaling dimension as $D_{\cal O}$ and
its free-field value as $D_{{\cal O},free}$.  The anomalous dimension
of this operator, denoted $\gamma_{\cal O}$, is defined via $D_{\cal
  O} = D_{{\cal O},free} - \gamma_{\cal O}$.  The coefficients $b_1$
and $b_2$ are independent of the scheme used for regularization and
renormalization and are $b_1=(1/3)[11C_A - 4T_fN_f]$
\cite{gw,politzer} and $b_2=(1/3)[34C_A^2 - 4(5C_A+3C_f)N_fT_f]$
\cite{caswell,jones}, where $C_2(R)$ is the quadratic Casimir
invariant, and $T(R)$ is the trace invariant, for the representation
$R$, and we use the notation $C_2(adj) \equiv C_A$, $T(R)\equiv T_f$,
and $C_2(R) \equiv C_f$. The AF condition means that $b_1 > 0$, i.e.,
$N_f < N_u$, where $N_u=11C_A/(4T_f)$.  Since $\alpha(\mu)$ is small
at large $\mu$, one can self-consistently calculate $\beta$ as a power
series in $\alpha(\mu)$.  As $\mu$ decreases from large values in the
ultraviolet (UV) to small values in the infrared (IR), $\alpha(\mu)$
increases.

A situation of special interest occurs if $\beta$ has a zero at a
nonzero (physical) value $\alpha = \alpha_{IR}$.  In the
asymptotically free regime, this happens if the condition
$N_u > N_f > 17C_A^2/[2(5C_A+3C_f)T_f]$ holds, 
so that $b_2 < 0$.  At the two-loop ($2\ell$) level, the zero in $\beta$
occurs at $\alpha_{IR,2\ell} = -4\pi b_1/b_2$. If $N_f$
is close enough to $N_u$ that this IR zero of $\beta$ occurs
at small enough coupling so that the gauge interaction does not
produce any spontaneous chiral symmetry breaking (S$\chi$SB), then it
is an exact IR fixed point (IRFP) of the RG.  The
theory at this IRFP exhibits scale invariance and is inferred to exhibit
conformal invariance, whence the term ``conformal window'' for this
regime.  In this IR limit, the theory is in a chirally symmetric,
deconfined, nonabelian Coulomb phase (NACP).  If, on the other hand,
as $\mu$ decreases and $\alpha(\mu)$ increases toward $\alpha_{IR}$,
there is a scale $\mu = \Lambda$ at which $\alpha(\mu)$ exceeds a
critical value, $\alpha_{cr}$, for the formation of a fermion
condensate $\langle \bar\psi\psi \rangle$ with associated S$\chi$SB,
then the fermions gain dynamical masses of order $\Lambda$. These fermions
are then integrated out of the low-energy effective
field theory operative for $\mu < \Lambda$.  In this case,
$\alpha_{IR}$ is only an approximate IRFP. We define $N_{f,cr}$ to be
the critical value of $N_f$ such that as $N_f$ decreases below $N_{f,cr}$,
there is S$\chi$SB.  If $N_f$ is only slightly less than $N_u$, so that
$\alpha_{IR}$ is small, then the theory at the IRFP is weakly coupled
and is amenable to perturbative analysis \cite{bz}. A case of
interest for studies of physics beyond the Standard Model (BSM) is
$N_f$ slightly less than $N_{f,cr}$. In this case, there is
slow-running, quasi-conformal behavior of $\alpha(\mu)$ over an
extended interval of $\mu$. The dynamical breaking of the approximate
scale (dilatation) symmetry then leads to a light
pseudo-Nambu-Goldstone boson, the dilaton.  In a BSM application, with
the Higgs boson being at least partially a dilaton, this might help to
solve the fine-tuning problem of why the Higgs mass is protected
against large radiative corrections.

It is of interest to investigate the properties of IRFPs in these
vectorial AF gauge
theories.  Among these properies are the anomalous dimensions of
(gauge-invariant) operators, such as $\bar \psi \psi =
\sum_{i=1}^{N_f} \bar\psi_i \psi_i$, denoted
$\gamma_{\bar\psi\psi,IR}$. In general, one can express the
anomalous dimension $\gamma_{\bar\psi/\psi}$  as the series expansion
\begin{equation}
\gamma_{\bar\psi\psi} = \sum_{\ell=1}^\infty c_\ell \, a^\ell \ , 
\label{gamma_series}
\end{equation}
where $c_\ell$ is the $\ell$-loop coefficient. Evaluating this with
$\alpha$ set equal to the IRFP value, calculated to a given $n$-loop
($n\ell$) order then yields $\gamma_{\bar\psi\psi,IR}$ to this order,
denoted as $\gamma_{\bar\psi\psi,IR,n\ell}$.  Another operator of
interest is ${\rm Tr}(F_{\lambda\rho}F^{\lambda\rho})$, where
$F^b_{\lambda\rho}$ is the field-strength tensor (with $b$ a group
index). The anomalous dimension $\gamma_{F^2,IR}$ of this operator at
the IRFP satisfies $\gamma_{F^2} = -\beta'_{IR}$, where $\beta' =
d\beta/d\alpha$.

As $N_f$ decreases through the conformal regime, $\alpha_{IR}$
increases, motivating higher-loop calculations of anomalous
dimensions.  We have carried out this program of calculating the UV to
IR renormalization-group evolution and anomalous dimensions at an IRFP
to higher-loop order in a series of papers, many with T. A. Ryttov,
including \cite{bvh}-\cite{dexm}.  Our first calculations were at the
4-loop level \cite{bvh}, and subsequently, we have extended these to
the 5-loop level, with inputs (in the $\overline{\rm MS}$ scheme) up
to the 5-loop level from \cite{b5su3,b5}. (At the 4-loop level, see
also \cite{ps}). Our calculations to higher-loop order enable us to
describe the IR properties of the theory throughout a larger portion
of the conformal window than would be possible with the lowest-order
(two-loop) results. As $N_f$ decreases below $N_{f,cr}$, the
properties of the IR theory change qualitatively, and the perturbative
calculations do not apply.  A unitarity upper bound in the conformal
regime is $\gamma_{\bar\psi\psi,IR} < 2$ (reviewed in
\cite{nakayama}), and studies of Schwinger-Dyson equations \cite{alm}
suggest that the onset of S$\chi$SB occurs if
$\gamma_{\bar\psi\psi,IR} > 1$. Thus, for a given $G$ and $R$, our
higher-loop calculations of $\gamma_{\bar\psi\psi,IR}$ yield estimates
for $N_{f,cr}$; in turn, this information is relevant for the
above-mentioned BSM theories.

There is an intensive ongoing program of research in the lattice gauge
theory community to study this physics.  Much work has been done for
$G={\rm SU}(3)$ with $R$ equal to the fundamental representation.
For this theory, $N_u=16.5$ (where a formal continuation from physical
integer $N_f$ to real $N_f$ is understood).  
There is not yet a consensus among lattice groups
concerning the value of $N_{f,cr}$ (i.e., the lower end of the conformal
window as a function of $N_f$) for this theory. As an example, we
consider the case $N_f=12$.  Several lattice groups
\cite{afn1,afn2,hasenfratz1,hasenfratz2,hasenfratz3,lombardo} have found  
that this theory is IR-conformal, while Ref. \cite{kuti}
has argued that it is chirally broken and hence not IR-conformal.
For our 5-loop analysis, we have made use of Pad\'e resummation methods
in addition to direct analysis of series expansions. 
As above, we denote our $n$-loop value of $\gamma_{\bar\psi\psi,IR}$ as
$\gamma_{\bar\psi\psi,IR,n\ell}$. We calculate
$\gamma_{\bar\psi\psi,IR,2\ell}=0.773$, 
$\gamma_{\bar\psi\psi,IR,3\ell}=0.312$, 
$\gamma_{\bar\psi\psi,IR,4\ell}=0.253$, and
$\gamma_{\bar\psi\psi,IR,5\ell}=0.255$. These results show reasonable
convergence at the 4-loop and 5-loop levels, and our values of
$\gamma_{\bar\psi\psi,IR,4\ell}$ and
$\gamma_{\bar\psi\psi,IR,5\ell}$ are in very good agreement 
with the values $\gamma_{\bar\psi\psi,IR} = 0.23(6)$
\cite{hasenfratz3} (in accord with \cite{hasenfratz1,hasenfratz2}) and
$\gamma_{\bar\psi\psi,IR} = 0.235(46)$ \cite{lombardo} measured in
lattice simulations. Our values are also
in agreement with the range of effective values reported in
\cite{kuti}.  For $\beta'_{IR}$ in this $N_f=12$ theory, as
calculated via power series in the IR coupling,
we find $\beta'_{IR,2\ell}=0.360$, $\beta'_{IR,3\ell}=0.295$, and
$\beta'_{IR,4\ell}=0.282$. Again, these values show good convergence, and
the 4-loop value is in very good agreement with the value
$\beta'_{IR}=0.26(2)$ obtained from lattice measurements
\cite{hasenfratz2}.
In our papers we have discussed corresponding comparisons
with lattice results for other gauge groups $G$, fermion representations $R$,
and flavor numbers $N_f$. We have also studied theories with fermions
in multiple different representations \cite{dexm}. 

Since the $b_\ell$ for $\ell \ge 3$ and the $c_\ell$ for $\ell \ge 2$ depend on
the scheme used for regularization and renormalization, it is
important to assess the effects of this scheme dependence.   We have
done this in \cite{sch,sch2,sch3,schl,schi}. This scheme dependence
is a generic feature of higher-order perturbative calculations, e.g., in
QCD. A scheme transformation can be
expressed as a mapping between $\alpha$ and $\alpha'$, or
equivalently, $a$ and $a'$, which we write as
$a = a' f(a')$, where $f(a')$ is the scheme transformation function.  We
can write $f(a')$ as a series expansion
\begin{equation}
f(a') = 1 + \sum_{s=1}^{s_{max}} k_s (a')^s  \ , 
\label{faprime}
\end{equation}
where $s_{max}$ may be finite or infinite.  In the new scheme, the
beta function is $\beta_{\alpha'} = -2\alpha' \sum_{\ell=1}^\infty
b'_\ell \, (a')^\ell$. We have calculated the $b'_\ell$ in terms of
the $b_\ell$ and $k_s$. In addition to the results $b_1'=b_1$ and
$b_2'=b_2$, we find
\begin{equation}
  b_3' = b_3 + k_1b_2 +(k_1^2-k_2)b_1 \ ,
    \label{b3p}
\end{equation}
\begin{equation}
  b_4'= b_4 + 2k_1b_3 + k_1^2b_2 +(-2k_1^3 + 4k_1k_2 - 2k_3)b_1 \ ,
\end{equation}
and so forth for higher $\ell$.  We have specified a set of conditions
that a physically acceptable scheme transformation must satisfy and
have shown that although these can easily be satisfied in the vicinity
of zero coupling, they are not automatic, and can be quite restrictive,
at a nonzero coupling, as is relevant for an IRFP in an UV-free (AF)
theory, or a UVFP in an IR-free theory.  As part of this work, we have
constructed scheme transformations that can map to a scheme with vanishing
coefficients at loop level $\ell \ge 3$ in the vicinity of the origin, but
we have also shown that it is more difficult to try to do this at a zero of
$\beta$ away from the origin.  We have applied these results to assess the
degree of scheme dependence in our higher-loop calculations of anomalous
dimensions at IRFPs in AF gauge theories and have shown that this dependence
is small.  This is similar to the experience in QCD, where calculations
performed to higher order exhibited reduced scheme dependence (e.g.
\cite{brodsky} and references therein).

The anomalous dimensions of gauge-invariant operators at the IRFP are physical
and hence cannot depend on the scheme used for regularization and
renormalization.  However, this property is not maintained by finite-order
perturbative series expansions beyond the lowest orders.  It is therefore
useful to calculate these anomalous dimensions in a scheme-independent
(SI) manner \cite{bz,grunberg,gtr}.  To do this, one utilizes the fact that
$\alpha_{IR} \to 0$ as $N_f \to N_u$. Hence, one can 
reexpress the anomalous dimensions as series
expansions in the manifestly scheme-independent variable $\Delta_f = N_u-N_f$,
rather than as power series in the IR coupling:
\begin{equation}
\gamma_{\bar\psi\psi,IR} = \sum_{j=1}^\infty \kappa_j \, \Delta_f^j
\label{gamma_ir_Deltaseries}
\end{equation}
and
\begin{equation}
\beta'_{IR} = \sum_{j=1}^\infty d_j \, \Delta_f^j \ ,
\label{betaprime_ir_Deltaseries}
\end{equation}
where $d_1=0$. In general, the calculation of the coefficient
$\kappa_j$ in Eq. (\ref{gamma_ir_Deltaseries}) requires, as inputs,
the values of the $b_\ell$ for $1 \le \ell \le j+1$ and the $c_\ell$
for $1 \le \ell \le j$.  The calculation of the coefficient $d_j$ in
Eq. (\ref{betaprime_ir_Deltaseries}) requires, as inputs, the values
of the $b_\ell$ for $1 \le \ell \le j$.  We denote the truncation of
these series to maximal power $j=p$ as
$\gamma_{\bar\psi\psi,IR,\Delta_f^p}$ and $\beta'_{IR,\Delta_f^p}$,
respectively.  With Ryttov we have calculated (i) the $\kappa_j$ up to
$j=4$, and thus the series expansion for $\gamma_{\bar\psi\psi,IR}$ to
$O(\Delta_f^4)$, and (ii) the $d_j$ up to $j=5$ and hence
$\beta'_{IR}$ to $O(\Delta_f^5)$ for general $G$ and $R$. 
We have studied a number of specific theories 
in detail, including the gauge groups SU($N_c$) with $R$ equal
to the fundamental, adjoint, and rank-2 symmetric and antisymmetric
tensor representations, and similarly for SO($N_c$) and Sp($N_c$) for
various $N_c$.  For the illustrative theory discussed above, namely
SU(3) with $N_f=12$ fermions in the fundamental representation, our
calculations of $\gamma_{\bar\psi\psi,IR}$ via
Eq. (\ref{gamma_ir_Deltaseries}) yield slightly larger values than our
calculations via Eq.  (\ref{gamma_series}), and our computations of
$\beta'_{IR}$ yield slightly smaller values than those that we
obtained via series expansions in the IR coupling.

An interesting feature of our scheme-independent results is that 
$\kappa_1$ and $\kappa_2$ are manifestly positive, and this
positivity also holds for $\kappa_3$ and $\kappa_4$ for a general $G$
and all of the representations $R$ that we have studied.  This leads
to two monotonicity properties in the conformal regime: (i) for a fixed $p$ with $1 \le p \le
4$, the anomalous dimension $\gamma_{\bar\psi\psi,IR,\Delta_f^p}$ is a
monotonically increasing function of $\Delta_f$, i.e., increases
monotonically with decreasing $N_f$; (ii) for a fixed
$N_f$, $\gamma_{\bar\psi\psi,IR,\Delta_f^p}$ is a monotonically
increasing function of $p$ in the range $1 \le p \le 4$.  From our
analyis of a ${\cal N}=1$ supersymmetric SU($N_c$) gauge theory with
$N_f$ conjugate pairs of chiral superfields \cite{dexss}, we have found that
this positivity property of the $\kappa_j$ is true for all $j$

\section{RG Studies of Other Theories}
\label{other_section}

We have also performed higher-loop studies of RG flows and possible
zeros of beta functions for other theories,
including (i) the 2D finite-$N$ Gross-Neveu model \cite{bgn},
(ii) various
$\phi^3$ theories in 6D \cite{phi36,rgb}, (iii) 4D U(1) gauge theory
\cite{lnf}, (iv) 4D
nonabelian gauge theories with $N_f > N_u$ \cite{lnf}, and (v) 4D O($N$) 
$\lambda |\vec \phi|^4$ theory \cite{lam1,lam2,lam3}.
The theories (i) and (ii) are
UV-free (i.e., AF), while the theories (iii)-(v) are IR-free. In these
studies, we combined direct analyses of higher-loop beta functions with
Pad\'e approximants and scheme transformations to derive results. 

\subsection{Finite-$N$ Gross-Neveu Model}

The Gross-Neveu (GN) model \cite{gn} is a 2D QFT with an $N$-component
massless fermion, $\psi_j$, $j=1,...,N$ and a four-fermion
interaction.  This model has been of interest because it exhibits some
properties similar to QCD, namely asymptotic freedom and formation of
massive bound states of fermions.  The model was solved exactly in the
$N \to \infty$ limit in \cite{gn}.  In this limit, the beta function
has no IR zero. This leaves open the question of whether the beta
function has an IR zero for finite $N$.  We investigated this in
\cite{bgn}, using the beta function up to the 4-loop level from
\cite{gls}. We found that, where the perturbative calculation of the
beta function is reliable, it does not exhibit robust evidence for an
IR zero.

\subsection{6D $\phi^3$ Theories}

$\phi^3$ theories in $d=6$ dimensions are asymptotically free, and it
is of interest to investigate whether they exhibit IRFPs.  We have
done this in \cite{phi36} with Gracey and Ryttov, using beta functions
calculated up to the 4-loop order.  As before, without loss of
generality, we take the matter field to be massless, since a $\phi$
field with nonzero mass $m_\phi$ would be integrated out of the
low-energy effective theory for momentum scales $\mu < m_\phi$ and
hence is not relevant for the IR limit $\mu \to 0$. We have studied
$\phi^3$ theories with a real 1-component $\phi$ field and also with
an $N$-component field $\phi_i$ transforming according to the
fundamental representation of a global SU($N$) symmetry, with a
self-interaction $\propto d_{ijk}\phi^i \phi^j \phi^k + h.c.$. For
both of these theories, we find evidence against an IRFP.
An interesting study of $\phi^3$ theory in a 6D spacetime with two
compact dimensions by Kisselev and Petrov is \cite{kp}.  In 
\cite{rgb}, we show that if a beta function is not
identically zero but has a vanishing one-loop term, then it is not, in
general, possible to use scheme transformations to eliminate
$\ell$-loop terms with $\ell \ge 3$ in the beta function, even in the
vicinity of the origin in coupling constant space.

\subsection{Studies of IR-free Theories, Including 4D U(1) and O($N$)
  $\lambda |{\vec \phi}|^4$}

If the $\beta$ function of a theory is positive near zero coupling, then this
theory is IR-free; as the reference scale $\mu$ decreases, the coupling
decreases toward 0. As $\mu$ increases from the IR, the coupling increases, and
a basic question is whether the beta function has a UV zero (in the
perturbatively calculable range), which would be a UV fixed point (UVFP)
of the RG.

\bigskip
\bigskip

An explicit example of a UVFP in an IR-free theory occurs in the
O($N$) nonlinear $\sigma$ model in $d=2+\epsilon$ dimensions. From
a solution of this model in the $N \to \infty$ limit, 
one finds, for small $\epsilon$ \cite{saclay,nlsm},
\begin{equation}
\beta(\xi) = \epsilon \xi \Big ( 1 - \frac{\xi}{\xi_c} \Big ) \ ,
\label{beta_nlsm}
\end{equation}
where $\xi$ is the effective coupling and $\xi_c = 2\pi \epsilon$.
Hence, assuming that $\xi$ is small for small $\mu$,
it follows that $\lim_{\mu \to \infty} \xi(\mu) = \xi_c$,
so the theory has a UVFP at $\xi_c$.

Let us consider a 4D U(1) gauge theory with $N_f$ fermions with a
charge $q$.  This theory is IR-free, and the 1-loop and 2-loop
coefficients in $\beta$ have the same sign, so there is no UV zero in
$\beta$ at the maximal scheme-independent order. 
In \cite{lnf} we investigated a possible UVFP at higher-loop order. 
One part of our work in \cite{lnf} was an analysis of the
beta function using the 5-loop coefficient
\cite{b5u1_kl,b5u1_bckr}. Another part made use of exact closed-form
results for $N_f \to \infty$ \cite{nfinf}.
In \cite{lnf} we also performed a corresponding
investigation of possible UVFP for a nonabelian gauge theory with $N_f
> N_u$.  In both the U(1) and nonabelian case, we found evidence
against a UVFP.  Of course, in neither case does this
imply that the theory has a
Landau pole, because the running gauge coupling gets
too large for perturbative calculations to be reliable before one
actually reaches this would-be pole.  

In \cite{lam1,lam2,lam3} we investigated the RG behavior of 4D O($N$)
$\lambda |{\vec \phi}|^4$ theory to six-loop order, using $b_5$ from
\cite{b5lam} and $b_6$ from \cite{b6lam} (in the $\overline{\rm MS}$
scheme).  Again, for values of the
interaction coupling where the perturbative (and Pad\'e resummation)
methods were applicable, we did not find robust evidence for a UVFP.

\section{Asymptotically Free Chiral Gauge Theories}

The analysis of asymptotically free chiral gauge theories is also of
considerable interest.  The (massless) fermion content is chosen so as to
avoid any gauge anomalies, mixed gauge-gravitational anomalies, and global
anomaly. As the theory flows from the UV to the IR and
the coupling grows, several possible types of behavior can occur,
including (i) an exact IRFP in a conformal phase; (ii) bilinear fermion
condensate formation with dynamical breaking of gauge and global
symmetries; or (iii) confinement with formation of massless composite
fermions. These theories have been of interest for BSM physics
(e.g, \cite{rds}). 
Our works in this area include \cite{df}-\cite{cgtakfb}, which
contain references to the extensive literature.

\section{Conclusion}

Studies of RG flows and possible RG fixed points in quantum field
theories continue to be of great interest, both from the point of
view of formal theory and for applications to BSM physics.
Here we have briefly discussed some of our results on 
higher-loop perturbative
calculations with inputs up to the five-loop level for anomalous
dimensions at IR fixed points in asymptotically free nonabelian gauge
theories and comparisons of these results with 
lattice measurements. We have also discussed our results on RG flows
and investigation of possible RG fixed points for several other UV-free
theories and for several IR-free theories.  There are many
opportunities for further work in this area.

\section*{Acknowledgements}

I am grateful to T. A. Ryttov and my other coauthors for fruitful
collaborations. I would also like to thank V. A. Petrov and the organizers
for kindly inviting me to give this talk at the XXXIII International Workshop 
on High-Energy Physics at the Logunov IHEP.

\paragraph{Funding information}
The research of R.S. was supported in part by the U.S. National Science
Foundation under the grants NSF-16-20628 and NSF-19-15093.


\nolinenumbers

\end{document}